\documentclass[apj,numberedappendix]{emulateapj}
\usepackage{apjfonts,graphicx,url,amssymb,amsmath,longtable,rotating,color,units,wasysym,subfigure,epsfig,listings,bm}
\usepackage{graphicx,amssymb,amsmath,epsfig}

\newcommand{\be}{\begin{eqnarray}}
\newcommand{\ee}{\end{eqnarray}}

\newcommand{\lp}{\left(}
\newcommand{\rp}{\right)}
\newcommand{\lb}{\left[}
\newcommand{\rb}{\right]}

\begin{document}

\slugcomment{Accepted for publication in The Astrophysical Journal}
\shorttitle{Constraints on Shallow $^{56}$Ni in SNe Ia}
\shortauthors{Piro, A. L. and Nakar, E.}

\normalsize


\title{Constraints on Shallow $^{56}$Ni from the Early Lightcurves of Type Ia Supernovae}

\author{Anthony L. Piro\altaffilmark{1} and Ehud Nakar\altaffilmark{2}}

\altaffiltext{1}{Theoretical Astrophysics, California Institute of Technology, 1200 E California Blvd., M/C 350-17, Pasadena, CA 91125; piro@caltech.edu}

\altaffiltext{2}{Raymond and Beverly Sackler School of Physics and Astronomy, Tel Aviv University, Tel Aviv 69978, Israel}


\begin{abstract}
Ongoing transient surveys are presenting an unprecedented account of
the rising lightcurves of Type Ia supernovae (SNe Ia). This early
emission probes the shallowest layers of the exploding white dwarf,
which can provide constraints on the progenitor star and the
properties of the explosive burning. We use semi-analytic models of
radioactively-powered rising lightcurves to analyze these
observations. As we have summarized in previous work, the main
limiting factor in determining the surface distribution of $^{56}$Ni
is the lack of an unambiguously identified time of explosion, as
would be provided by detection of shock breakout or shock-heated
cooling. Without this the SN may in principle exhibit a ``dark
phase'' for a few hours to days, where the only emission is from
shock-heated cooling that is too dim to be detected.
We show that by assuming a theoretically-motivated time-dependent
velocity evolution, the explosion time can be better constrained, albeit
with potential systematic uncertainties. This technique is used to infer the surface $^{56}$Ni distributions of three recent SNe~Ia that were caught especially early in their rise. In all three we find fairly similar $^{56}$Ni distributions. Observations of SN 2011fe and SN 2012cg
probe shallower depths than SN 2009ig, and in these two cases $^{56}$Ni is present merely $\sim10^{-2}M_\odot$ from the WDs' surfaces. The uncertainty in this result is up to an order of magnitude given the difficulty of precisely constraining the explosion time. We also use our conclusions about the explosion times to reassess radius constraints for the progenitor of SN~2011fe, as well as discuss the roughly $t^2$ power law that is inferred for many observed rising lightcurves.
\end{abstract}



\section{Introduction}
\label{sec:introduction}

Type Ia supernovae (SNe Ia) play a central role in
modern astrophysics. They are used as distance
indicators to probe the expansion of the Universe
\citep{rie98,per99}, they produce most of the iron-group elements in
the cosmos \citep{Iwamotoetal1999}, and they provide an astrophysical context for studying
explosions \citep{HillebrandtNiemeyer2000}. But their importance has brought attention to
the theoretical uncertainties that frustratingly remain. It is
generally accepted that they result from unstable thermonuclear
ignition of degenerate matter \citep{hf60} in a C/O white dwarf
(WD), but the progenitor systems have not been identified.
Candidates include stable accretion from a non-degenerate binary
companion \citep{wi73}, the merging of two C/O WDs
\citep{it84,web84}, or accretion and detonation of a helium shell on
a C/O WD that leads to core detonation \citep{ww94a,la95}. In
addition, it is not known whether the incineration proceeds
as a sub-sonic deflagration \citep{nom76,nom84} or
deflagration-detonation transition \citep[DDT;][]{kho91,ww94b}. Each of these
scenario has implications for the velocity profile,
density structure, and distribution of ashes within the exploding
WD.

A powerful method for constraining between these models is the
study of the early-time behavior of SNe Ia, since this is
when the shallowest layers of the WD are probed by the observed emission. Analysis of spectra
provides one way of learning about the surface abundances of these
explosions \citep[e.g.,][]{Hachingeretal2013}. The photometry is also
sensitive to the depth and distribution of radioactive heating
\citep{Piro2012,PiroNakar2013}. With early observations of SNe Ia
becoming more common, the time is ripe to explore what can be
learned from these measurements.

In the following work we use semi-analytic models to study where and
how much $^{56}$Ni is present in the outer ejecta of SNe Ia. As
discussed in our previous investigation of radioactively-powered
lightcurves \citep{PiroNakar2013}, it is difficult to directly
measure the $^{56}$Ni distribution without a detection of the
explosion time, as would be provided by shock breakout or
shock-heated cooling
\citep{Piroetal2010,NakarSari2010,NakarSari12,Rabinaketal2012}.
Unfortunately, in the case of SNe Ia, such emission has never been
detected because of the small WD radius. If merely photometric
lightcurves of the rise are available, there is a degeneracy between emission
being from $^{56}$Ni near the surface with a recent explosion versus
$^{56}$Ni deeper in the star but with an explosion further in the
past. In the latter case, a SN Ia exhibits a ``dark phase'' for a
few hours to days until the thermal diffusion wave reaches the
shallowest $^{56}$Ni deposits. Even with these uncertainties,
constraints can still be provided by comparing a wider range of
properties, such as the velocity evolution. This
information is available for a few well-studied SNe Ia, and we use
it in order to estimate the time of explosion and surface $^{56}$Ni
distribution for each of them.

In \S \ref{sec:theory} we summarize the semi-analytic framework used to model the rising lightcurves. In \S \ref{sec:comparison} we analyze observations of three recent SNe~Ia and summarize our constraints on their shallow $^{56}$Ni distributions. In \S \ref{sec:t^2} we consider the $t^2$ rise that is often observed in early lightcurves and discuss whether
$t^2$ (or any power law) should be expected. We conclude in \S \ref{sec:conclusion} with a summary of our results and a discussion of potential future work.

\section{Radioactively-Powered Rising Lightcurves}
\label{sec:theory}

In the following we present the model used for this study, which
borrows from and builds upon our recent work on
radioactively-powered rising lightcurves. In \citet{Piro2012}, we
focused on direct $^{56}$Ni heating at the depth of the diffusion
wave. In \citet{PiroNakar2013}, we added the ``diffusive tail,''
which provides heating at depths shallower than the intrinsic
$^{56}$Ni distribution. Here we include these effects in greater
detail by integrating over their contributions throughout the WD, as
discussed in Appendix~B of \citet{PiroNakar2013}. We quickly summarize
the main results here for completeness.

As the ejecta from the SN expands, a thermal diffusion wave travels back through the material. This is defined as the depth at which photons can diffuse up to the surface of the exploding star within the time since the start of the explosion. This condition is satisfied where the optical depth to the observer is approximately $c/v$, where $c$ is the speed of light and $v$ is the velocity of the expanding gas at the location of the diffusion wave. Note that the diffusion depth (at optical depth greater than unity) is considerably deeper than the photosphere. At any time $t$, the diffusion wave has a
depth of roughly
\be\label{eq:mdiff}
    \Delta M_{\rm diff}\approx 2\times10^{-2}\frac{E_{51}^{0.44}}{\kappa_{0.1}^{0.88}M_{1.4}^{0.32}} \lp\frac{t}{1\,{\rm day}} \rp^{1.76}M_\odot,
\ee
where $E=10^{51}E_{51}\,{\rm erg}$ is the explosion energy, and $M=1.4M_{1.4}\,M_\odot$ is the ejecta mass, and $\kappa=0.1\kappa_{0.1}\,{\rm cm^2\ g^{-1}}$ is the opacity. We approximate the opacity as constant. This is motivated by the fact that during the times at which we are modeling these events the bolometric luminosity is always greater than $10^{41}\,{\rm erg\,s^{-1}}$. Combined with the times of explosion that we derive, along with the observed photospheric velocities, we infer that the temperature at the diffusion depth is always $> 10,000\,{\rm K}$, during the rising phase. Thus, carbon and oxygen are always ionized, at least once, and if these elements dominate the opacity then it is in the range $0.03-0.2\,{\rm cm^2\,g^{-1}}$. If the outer layers have sufficient $^{56}$Ni to dominate the opacity, then it is $\sim 0.1\,{\rm cm^2\,g^{-1}}$ \citep{PintoEastman2000}. Therefore, our opacity assumption introduces at most a factor of 3 error in the diffusion depth. The scalings and prefactors in equation (\ref{eq:mdiff}) use Appendix C of \citet{PiroNakar2013} with values appropriate for Chandrasekhar mass WDs.

At the times we consider, the ejecta is optically thick to gamma-rays emitted from radioactive decay, and they efficiently heat the SN. Heating in material shallower than $\Delta M_{\rm diff}$ directly goes into the observed luminosity. Heating in material deeper than $\Delta M_{\rm diff}$ only contributes to the observed lightcurve if some fraction of the photons from these larger depths are able to diffuse up to $\Delta M_{\rm diff}$. This produces the so-called diffusive tail. Motivated by this picture, we split the total observed luminosity into two parts
\be
    L(t) = L_{\rm direct}(t) + L_{\rm tail}(t),
\ee
where $L_{\rm direct}$ is the direct heating by $^{56}$Ni down to $\Delta M_{\rm diff}$, and $L_{\rm tail}$ is the diffusive tail from material deeper than $\Delta M_{\rm diff}$. Each is an integral over different regions of the ejecta. For the direct heating component
\be\label{eq:l56}
    L_{\rm direct}(t) = \int^{t}_{0} X_{56}(t')\frac{\partial \Delta M_{\rm diff}}{\partial t'} \epsilon(t) dt',
\ee
where $X_{56}(t)$ is the mass fraction of $^{56}$Ni at the depth of the diffusion wave at time $t$, and the specific heating rate is
\be
    \epsilon(t) = \epsilon_{\rm Ni} e^{-t/t_{\rm Ni}} + \epsilon_{\rm Co} ( e^{-t/t_{\rm Co}}- e^{-t/t_{\rm Ni}} ),
    \label{eq:e56}
\ee
where $\epsilon_{\rm Ni}=3.9\times10^{10}\,{\rm erg\ g^{-1}\ s^{-1}}$, $t_{\rm Ni}=8.76\,{\rm days}$, $\epsilon_{\rm Co}=7.0\times10^9\,{\rm erg\ g^{-1}\ s^{-1}}$, and $t_{\rm Co}=111.5\,{\rm days}$. The total diffusive tail component is the integral over all the diffusive tails from heating deeper than $\Delta M_{\rm diff}$,
\be\label{eq:ltail}
    L_{\rm tail}(t)=
    \int^{t_{\rm diff}}_{t} X_{56}(t')\frac{\partial \Delta M_{\rm diff}}{\partial t'}
    \epsilon(t) \frac{{\rm erfc}( t'/\sqrt{2}t )}{ {\rm erfc}(1/\sqrt{2} )} dt'.
\ee
We take the upper integration limit to be the diffusion time through the entire ejecta $t_{\rm diff}$\footnote{Note that in this work we are using a different definition of $t_{\rm diff}$ than that in \citet{PiroNakar2013}.}, which roughly corresponds to the time of lightcurve peak.

Since $\Delta M_{\rm diff}\propto t^{1.76}$, equations (\ref{eq:l56}) and (\ref{eq:ltail}) are rewritten as
\be
     L_{\rm direct}(t)= 1.76 L_{56}(t)
        \int^{t}_{0}
    \frac{X_{56}(t')}{X_{56}(t)}
    \left(\frac{t'}{t}\right)^{1.76}
    \frac{dt'}{t'},
\ee
and
\be
        L_{\rm tail}(t)=1.76 L_{56}(t)
        \int^{t_{\rm diff}}_{t}
        \frac{X_{56}(t')}{X_{56}(t)}
        \left(\frac{t'}{t}\right)^{1.76}
        \frac{{\rm erfc}( t'/\sqrt{2}t )}{ {\rm erfc}(1/\sqrt{2} )}
        \frac{dt'}{t'},
        \nonumber
        \\
\ee
where
\be\label{eq:luminosity_local}
	L_{56}(t)\equiv X_{56}(t)\Delta M_{\rm diff}(t)\epsilon(t),
\ee
is roughly the local heating rate from $^{56}$Ni. The luminosity has no contribution from the diffusive tail once the diffusion wave has travelled through the ejecta, thus we define $L_{\rm diff}\equiv L_{\rm direct}(t=t_{\rm diff})$.

When actually performing calculations, it is useful to write these expressions in dimensionless forms. First, let $x\equiv t/t_{\rm diff}$ and $x'\equiv t'/t_{\rm diff}$, where $x$ and $x'$ vary from $0$ to $1$. We define the ratio of the local heating rate to $L_{\rm diff}$ as
\be\label{eq:lambda}
    \Lambda(x)
    &\equiv &1.76\frac{L_{56}(x)}{L_{\rm diff}}
    \nonumber
    \\
    &=& \frac{\epsilon(x)}{\epsilon(1)}
    \lb \int^{1}_{0}
    \frac{X_{56}(x')}{X_{56}(x)}
    \lp \frac{x'}{x}\rp^{1.76}\frac{dx'}{x'} \rb^{-1}.
\ee
The ratio of the observed time-dependent luminosity to $L_{\rm diff}$ is then
\be\label{eq:luminosity_theory}
       \frac{L(x)}{L_{\rm diff}}
        &=&
        \Lambda(x)
    \int^{x}_{0} \frac{X_{56}(x')}{X_{56}(x)} \left(\frac{x'}{x}\right)^{1.76} \frac{dx'}{x'}
    \nonumber
    \\
    &+&
     \Lambda(x)
    \int^{1}_{x} \frac{X_{56}(x')}{X_{56}(x)} \left(\frac{x'}{x}\right)^{1.76}
        \frac{{\rm erfc}( x'/\sqrt{2}x )}{ {\rm erfc}(1/\sqrt{2} )} \frac{dx'}{x'}.
\ee
In this form the right-hand side is dimensionless and only depends on the $^{56}$Ni distribution. This allows us to vary $X_{56}(x)$ and calculate a wide range of lightcurves, which can then be rescaled to a particular observation via $L_{\rm diff}$ and $t_{\rm diff}$.

When fitting a $^{56}$Ni distribution to a given lightcurve in the next section, we use the parametrization
\be\label{eq:x56}
    X_{56}(x) = \frac{X_{56}'}{1+\exp\lb-\beta(x-x_{1/2}) \rb},
\ee where $X_{56}'$ sets the normalization, $\beta$ controls the
steepness of the rise, and $x_{1/2}$ is the time when
$X_{56}/X_{56}'= 1/2$. This allows us to consider a variety of
$^{56}$Ni distributions with two parameters. The normalization is
determined by
\be
    X_{56}' &=& \frac{L_{\rm diff}}{1.76\Delta M_{\rm diff}(t_{\rm diff})\epsilon(t_{\rm diff})}
    \nonumber
    \\
    &&\times\lb \int^{1}_{0}
    \frac{x'^{0.76}dx'}{1+\exp\lb-\beta(x'-x_{1/2}) \rb}
     \rb^{-1},
\ee
and thus is not a free parameter. The drawback of this parameterization is that we can only consider $^{56}$Ni distributions that increase with depth. A more complicated distribution is a realistic possibility, such as in a double detonation where there may be a surface enhancement of $^{56}$Ni from explosive burning of a helium shell \citep{sb09,fin10}. In future studies we will better explore such $^{56}$Ni distributions.

\section{Comparisons to Specific Supernovae}
\label{sec:comparison}

Recent observations have been especially fruitful in catching SNe Ia at early times. We use this work to analyze three well-studied events: SN~2011fe, SN~2012cg, and SN~2009ig. For each we summarize what can be constrained from their photometric lightcurves and velocity evolution. Although there are particular issues for each event (which we discuss below), our general strategy is as follows.
\begin{enumerate}
\item Since a SN may in principle exhibit a dark phase, we assume that the time of explosion is not known.
\item For a spectral line generated at constant specific opacity, its velocity is a power law with time with $v\propto t^{-0.22}$ \citep{PiroNakar2013}.
We vary the explosion time and check when the observed absorption
features best match this power law. From this we infer what is
the likely explosion time.
\item The photospheric velocity $v_{\rm ph}$ is expected to roughly follow the low-velocity Si II $\lambda6355$ absorption feature \citep{Tanakaetal2008}. Using the fits performed in the previous step, we can therefore estimate $v_{\rm ph}(t)$. The photospheric radius is then given by $r_{\rm ph} = v_{\rm ph}t$.
\item Assuming that the SN emits roughly as a blackbody and using the observed $B$, $V$, and $R$ lightcurves, we fit the color temperature $T_c$ and bolometric luminosity as a function of time using $L\approx 4\pi r_{\rm ph}^2\sigma_{\rm SB}T_c^4$. Using just these wave bands, the inferred bolometric luminosity is always a lower limit.
\item Theoretical lightcurves are generated with different $X_{56}(x)$ via equation (\ref{eq:luminosity_theory}), where $X_{56}(x)$ has the functional form of equation (\ref{eq:x56}). We estimate $L_{\rm diff}$ and $t_{\rm diff}$ as roughly the peak luminosity and time of peak luminosity, respectively. In this way the theoretical lightcurves are rescaled for comparison with the bolometric lightcurve, and we can put constraints on what is the most likely distribution of $^{56}$Ni.
\end{enumerate}
The largest limitation of this framework is our assumption of a specific time-dependent power law for the velocity evolution of $v\propto t^{-0.22}$ in step 2 above. As we show below, we find that all three of the absorption features we focus on roughly obey this same power-law dependence\footnote{Interestingly, this indicates that these different features are due to
different line opacities within a flow with the same velocity
power-law profile and are not separate velocity components in the
ejecta. This argues against situations where the high velocity
features are generated by a separate event during or prior to the
explosion \citep[e.g.,][]{Piro2011}.}. This would not have been the case if the method we use is entirely wrong, and thus this lends some support for our approach. Nevertheless, it is possible that small variations from the theoretically predicted power law introduce systematic errors. We try to better quantify the errors introduced by this fundamental assumption of our model by varying the exponent of the power law from $0.20-0.24$ (see the discussions in the following sections). This shows that it is difficult to constrain the explosion time to better than roughly $\pm0.5\,{\rm days}$. Later this rough error is also used to quantify the uncertainty in the derived $^{56}$Ni distributions. Beyond this, it is difficult for us to further quantify how much different the $^{56}$Ni distribution could be if, for example, the photosphere evolved in a much more complicated way. A useful exercise would therefore be to use detailed numerical modeling from explosion simulations to better test these assumptions.

\subsection{Modeling SN 2011fe}

We first focus on SN 2011fe because it is the most constrained by our modeling. SN 2011fe exploded in August 2011 as the closest SNe Ia in the last 25 years \citep{Nugentetal2011}. The considerable interest in this event and its proximity make it one of the best studied SNe Ia. The time-dependent velocities of absorption features are summarized in \citet{Parrentetal2012}.
The $B$, $V$, and $R$ rising lightcurves are presented in
\citet{Vinkoetal2012}. This particular work was chosen because of the high
density of observations during the rise, but we could have just as
well considered other data sets
\citep{RichmondSmith2012,Munarietal2013}. We use a distance modulus
for M101 of $29.05$ \citep{ShappeeStanek2011}, and no reddening is
included because it has been inferred to be relatively small
\citep{Patatetal2013}. The various studies have found different times for the
peak bolometric luminosity, depending on the fitting method used. For the present work we choose a time of peak of ${\rm
JD}~2455815.4$, although this choice does not greatly impact our conclusions for the $^{56}$Ni distribution at shallow
depths. The earliest detection was at ${\rm
JD}~2455797.65$ \citep{Nugentetal2011}, and their fitting of a $t^2$ power law to the rising luminosity,
(as is common practice) gives an explosion time of ${\rm
JD}~2455797.2$ before the peak. Although they quote an error of $\pm0.01\,{\rm days}$, as we discuss later this
practice is not well justified and the true uncertainty in the
explosion time is considerably larger. \citet{Nugentetal2011} also observed the location of SN 2011fe roughly at ${\rm JD}~2455796.7$, which provides an upper limit in the apparent $g$-band magnitude of $21.5$ (absolute magnitude of $-7.55$).

\begin{figure}
\epsscale{1.2}
\plotone{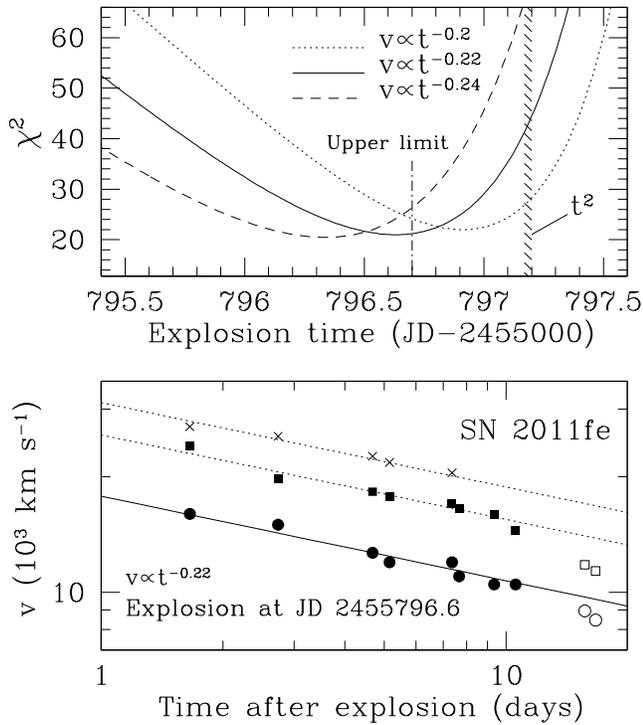}
\caption{The top panel plots the $\chi^2$ found by fitting the velocity evolution as a power law with time for different explosion times and various power-law indices as labeled. We draw a vertical dot-dashed line at the time of an upper limit from a non-detection and a shaded region at the explosion time inferred by fitting a $t^2$ rise \citep{Nugentetal2011}. In the bottom panel  we plot the observed low-velocity Si II $\lambda6355$ (circles), high-velocity Si II $\lambda6355$ (squares), and high-velocity Ca II H\&K (crosses). Filled and open symbols indicate data that was used or not used for the fit, respectively. The lines show our best fit velocity evolution for $v\propto t^{-0.22}$, and the solid line indicates the $v_{\rm ph}$ we use in subsequent analysis.}
\label{fig:2011fe_velocity}
\epsscale{1.0}
\end{figure}

In our analysis of the velocity evolution, we use the low-velocity
Si II $\lambda6355$, high-velocity Si II $\lambda6355$, and
high-velocity Ca II H\&K absorption features\footnote{Other
absorption features are measured, but we restrict our study to these
three since they are some of the most widely available in SN Ia
literature.}. The velocities of the absorption lines are always calculated from the location of deepest absorption.
In the top panel of Figure \ref{fig:2011fe_velocity}
we plot the $\chi^2$ found by fitting these features with power-law
velocity profiles as a function of the explosion time, where $\chi^2$ is
defined as
\be
    \chi^2 = \sum\limits_N \lp \frac{v_N-v(t)}{\Delta v}\rp^2,
\ee
where $N$ is the number of data points, $v_N$ is a measured velocity,
and $\Delta v=500\ {\rm km\ s^{-1}}$ is a rough estimate of the
measurement error \citep{Parrentetal2012}. The reduced $\chi^2$ around the best fit
explosion time is about 1.2 (there are 17 degrees of freedom). In Figure \ref{fig:2011fe_velocity} we
consider three different
velocity power-law indices centered around the model prediction of
$v\propto t^{-0.22}$. This shows
that the model provides a good description of the data, and that
assuming that the power-law index is known, the explosion time is
measured to within about $\pm 0.25$ days. However,  assuming slightly different power laws produces fits with
similar quality and results in explosion times that vary by about $\approx1$ day. Since
theoretically $v\propto t^{-0.22}$ is the preferred velocity profile
we consider ${\rm JD}~2455796.6$ to be the most
likely explosion time with an uncertainty of roughly $\pm 0.5\,{\rm day}$. This is actually very similar (within $0.1\,{\rm days}$) of the non-detection by \citet{Nugentetal2011}. In the bottom
panel we present the velocity data along with our best-fit velocity
evolutions. Open symbols indicate data that were not used for the
fit because they are near peak where the velocity profile
is not expected to be a power law.

\begin{figure}
\epsscale{1.2} \plotone{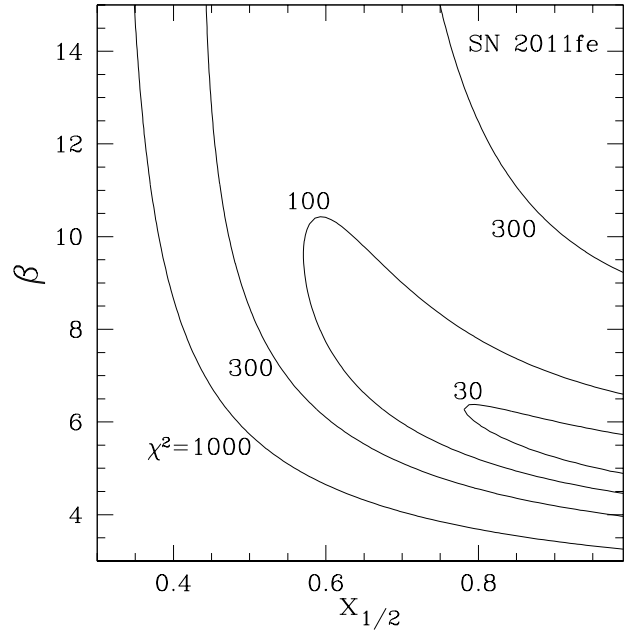}
\caption{Contours of constant $\chi^2$ (as labeled) from fitting for the $^{56}$Ni distribution (through $\beta$ and $x_{1/2}$) needed to explain the rising lightcurve of SN 2011fe. This demonstrates that a relatively low value of $\beta$ is needed, which in turn implies a shallow distribution of $^{56}$Ni. Although not presented in this paper, the fits for SNe 2009ig and 2012cg are similar.}
\label{fig:chisq_11fe} \epsscale{1.0}
\end{figure}

\begin{figure}
\epsscale{1.2} \plotone{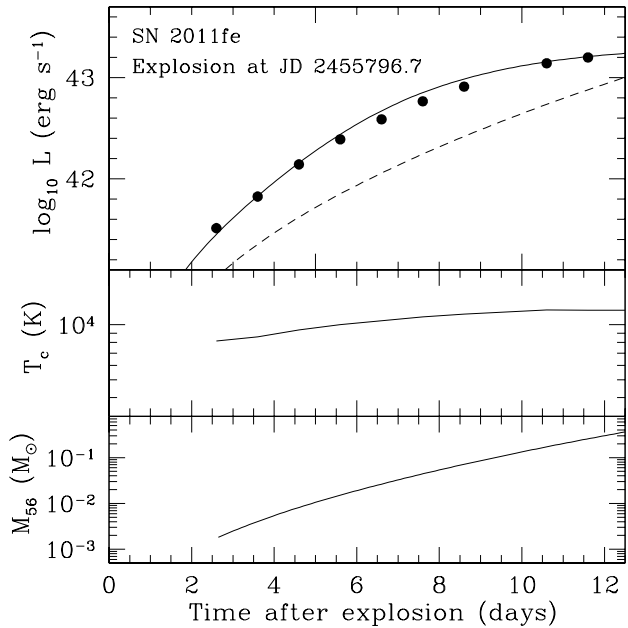} \caption{A summary of
the fits to SN 2011fe. The top
panel shows the inferred bolometric luminosity (filled circles), the
fit bolometric luminosity (solid curve), and the local heating from
$^{56}$Ni of $L_{56}$ (dashed curve) given by equation (\ref{eq:luminosity_local}). The
middle panel shows the color temperature, and the bottom panel shows
$M_{56}=L_{56}/\epsilon$.} \label{fig:2011fe} \epsscale{1.0}
\end{figure}

For any given explosion time we can look for the $^{56}$Ni distribution that
produces the observed luminosity. The fitting is done via a $\chi^2$ minimization
over $\beta$ and $x_{1/2}$ in the parameterization of $X_{56}$ given by
equation (\ref{eq:x56}). Assuming a $\approx10\%$ error in the bolometric
luminosity measurements, the $\chi^2$ per degree of freedom of the best
fit $^{56}$Ni distribution is less than two. Contours of constant $\chi^2$ are plotted in
Figure \ref{fig:chisq_11fe} to demonstrate the quality of the fit and how much degeneracy
there is. Although this does not prove that
the $^{56}$Ni distribution we derive is unique, it at least shows that it
does a good job of modeling the data. The results from fitting the
photometric observations are presented in Figure \ref{fig:2011fe}. In this
particular case we use the time of the non-detection for the explosion time, which is sufficiently
close to our preferred time so that the qualitative features are unchanged.
In the top panel we
compare the inferred bolometric lightcurve (filled circles) to the
model fit (solid curve).  We also plot the contributions from local
heating $L_{56}$ (dashed curve) to show how well the bolometric luminosity
reflects the underlying $^{56}$Ni distribution. We find a range of $L_{56}/L\sim 0.2-0.6$,
and typically $L_{56}/L\sim 0.3$, during the early times of the SNe.
This means that although the match is not exact, the underlying $^{56}$Ni distribution
is roughly represented by the observed luminosity and nonlocal effects are not dominating.
Therefore $^{56}$Ni must be present, at least in some amount, at the depths that are probed by
the earliest emission. To test the robustness of this conclusion, we
varied the $^{56}$Ni distribution (by varying $\beta$ and $x_{1/2}$) by
two standard deviations from the best-fit values. We still found that $^{56}$Ni
must be present near the exploding star's surface, showing that it is difficult to
explain the early rise without some $^{56}$Ni at the diffusion depth.

The middle panel of Figure \ref{fig:2011fe} shows the
inferred color temperature $T_c$. This confirms our earlier
discussion of the opacities in \S \ref{sec:theory}, and that the temperature at
the diffusion depth (which is greater than $T_c$ by
a factor of $\approx\tau^{1/4}$, where $\tau \approx 30$ is the optical
depth at the diffusion depth) is always sufficiently high that
carbon and oxygen will not be fully recombined.

The bottom panel shows the
mass of $^{56}$Ni above the diffusion wave depth, given by
\be
    M_{56}(t)=L_{56}(t)/\epsilon(t).
\ee
This is roughly independent of the explosion time because it is just
set by the bolometric luminosity at any given time. In contrast,
$T_c$ changes with explosion time because an explosion further in the past
implies more expansion at any given time and thus a smaller
$T_c$. This means that an additional constraint on the explosion
time could be made via a temperature measurement, although this
requires detailed spectral modeling that is outside the scope of
this work \citep[see the discussion of $t_{\rm min}$
in][]{PiroNakar2013}.

\subsection{Radius Constraints and Shallowest $^{56}$Ni for SN
2011fe}\label{sec:2010fe radius}

Using the data from \citet{Nugentetal2011} and a
non-detection $\approx7\,{\rm hrs}$ earlier, \citet{Bloometal2012} argued that the progenitor of
SN 2011fe had a radius $\lesssim0.02R_\odot$ by using models of
shock-heated cooling \citep{Piroetal2010,Rabinaketal2012}. But this
assumed that the time of explosion could be accurately determined
from extrapolating $t^2$ back in time. As emphasized in
\citet{PiroNakar2013}, this is not generally a robust method for
finding the explosion time (see also \S \ref{sec:t^2}),
so it is worth revisiting the radius constraint for
a range of explosion times.

\begin{figure}
\epsscale{1.2}
\plotone{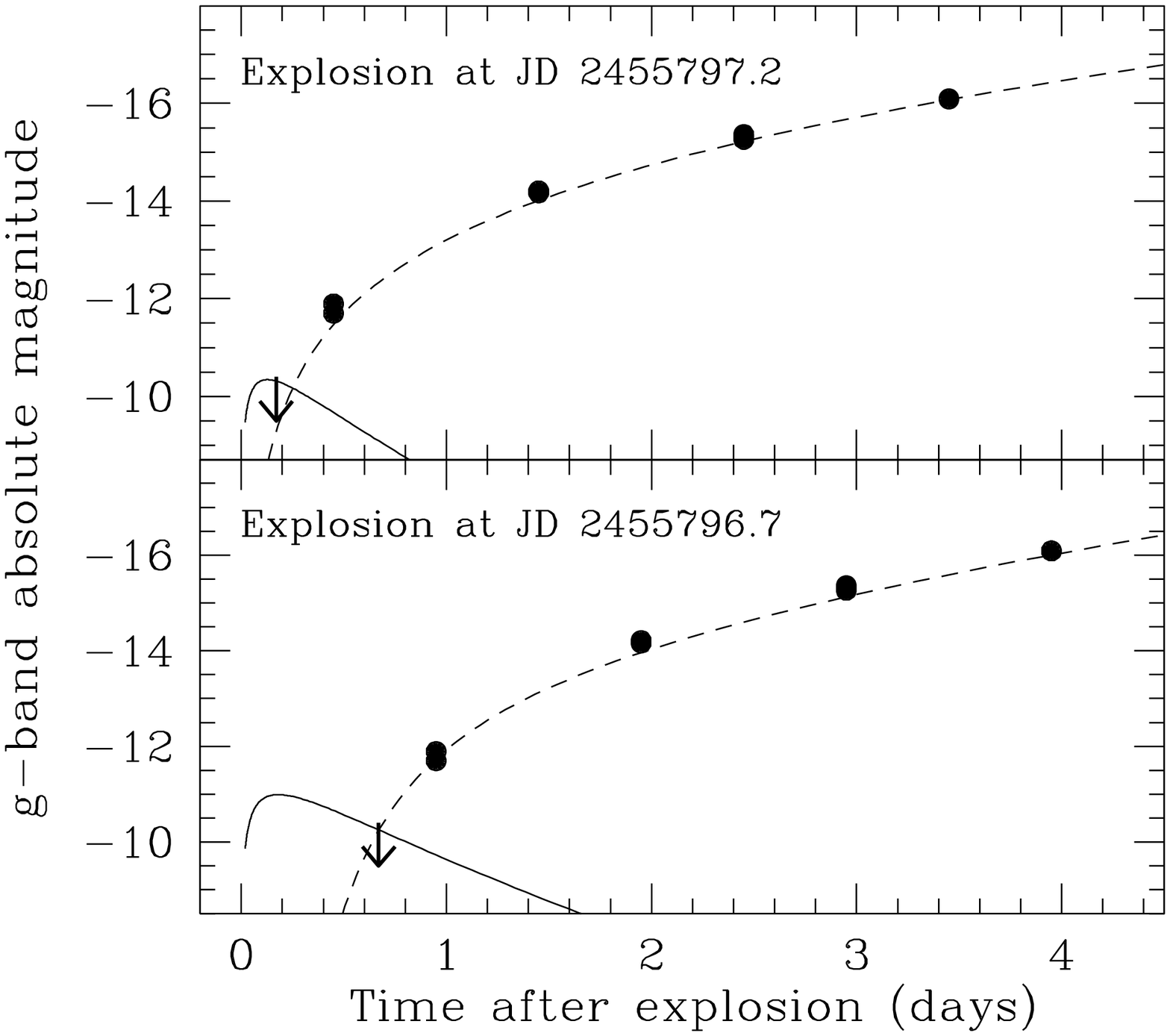}
\caption{Comparison of the early $g$-band data \citep{Nugentetal2011} and a non-detection upper limit \citep{Bloometal2012} to theoretical lightcurves from radioactive heating (dashed curves) and shock-heated cooling (solid curves) calculated according to \citet{Piroetal2010}. This does not include the suppression of the shock-heated cooling (or ``drop out'') that occurs when the diffusion wave moves into ideal gas dominated material \citep{Rabinaketal2012}. The top panel is roughly the explosion time inferred from a $t^2$ extrapolation. The bottom panel assumes that the explosion occurred $0.5\,{\rm days}$ earlier, for which the radius constraint is a factor of $1.9$ larger.}
\label{fig:radius}
\epsscale{1.0}
\end{figure}

In Figure \ref{fig:radius} we plot the early data and non-detection
upper limit for SN 2011fe for two different explosion times. The theoretical curves include
radioactive heating (dashed curves) and shock-heated cooling (solid
curves). The first thing to note is that $^{56}$Ni cannot always be
present at the earliest times and still produce the observed
lightcurves. In the bottom panel we had to cut off the $^{56}$Ni for times earlier than $0.9\,{\rm days}$ after explosion
in order to not overpredict the $g$-band upper limit reported in
\citet{Bloometal2012}. (In the top panel no $^{56}$Ni cut-off is needed.)
This implies that for earlier explosion times there is a sharp cut-off in the $^{56}$Ni distribution near the depth that generates the
luminosity of the first detected light. This is not unexpected since $^{56}$Ni probably does not extend to the
very surface and the earliest emission will be due to the diffusive
tail. In \S \ref{sec:compare} we further discuss what depth in the
WD is implied by this time.

The other thing to note from Figure \ref{fig:radius} is that when
the explosion time is further in the past, upper limits on the emission from
shock-heated cooling (solid curves) are not as stringent. Using the models
from \citet{Piroetal2010} we find that when the explosion is merely $0.5\,{\rm days}$
further in the past (the bottom panel) the radius can be a factor of $1.9$ greater than in the top
panel.

Another potentially important effect that is not included in Figure \ref{fig:radius} is the ``drop out''
in the shock-heated cooling emission that is expected once the diffusion
wave exposes the depth where the shock is matter rather than radiation dominated. This
is expected to occur $\sim\,{\rm hours}$ after explosion for a typical WD radius \citep{Rabinaketal2012}.
Although we do not consider explosion times earlier than the non-detection of \citet{Nugentetal2011} in
Figure \ref{fig:radius}, it is possible that the explosion occurred before ($\sim0.5\,{\rm day}$ ) this time, and
the non-detection is simply during the dark phase between the drop out and the latter $^{56}$Ni heating.
\citet{Bloometal2012} find that the drop out limits the radius constraint posed by their upper limit.
If the explosion is one day before the date estimated by \citet{Nugentetal2011}, then the uncertainty in the limit on the radius
of the progenitor of SN 2010fe is somewhat increased. Hence the progenitor can be as large as $\sim0.1R_\odot$.

\subsection{SN 2012cg}

\begin{figure}
\epsscale{1.2}
\plotone{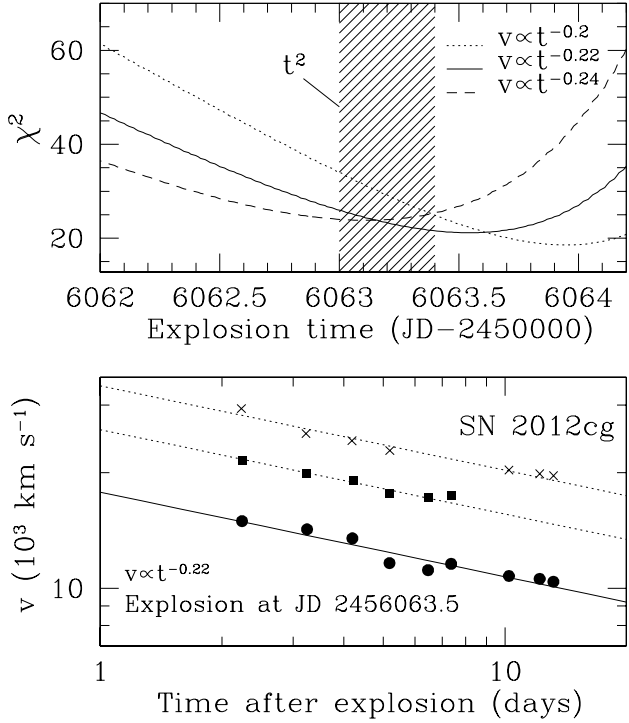}
\caption{The same as Figure \ref{fig:2011fe_velocity}, but for SN 2012cg. The shaded region shows the inferred explosion time from \citet{Silvermanetal2012} using $t^2$.}
\label{fig:2012cg_velocity}
\epsscale{1.0}
\end{figure}

\begin{figure}
\epsscale{1.2} \plotone{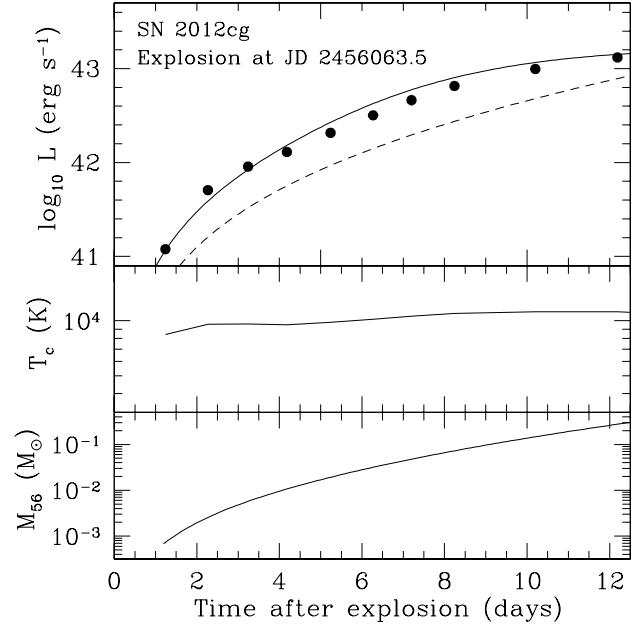} \caption{The same as
Figure \ref{fig:2011fe}, but for SN 2012cg.} \label{fig:2012cg} \epsscale{1.0}
\end{figure}

The velocities and photometry for SN 2012cg are summarized in
\citet{Silvermanetal2012}. Further photometry is presented by \citet{Munarietal2013},
including data around the peak identified to occur at roughly ${\rm JD}~2456083.0$. The velocity fitting results are shown
Figure \ref{fig:2012cg_velocity} (again taking \mbox{$\Delta v=500\ {\rm km\ s^{-1}}$}). The best fit explosion time is ${\rm JD}~2456063.5$, but the strength of the fit is not as
strong as for SN 2011fe. In comparison, using $t^2$
\citet{Silvermanetal2012} find ${\rm JD}~2456063.2\pm0.2$ (the shaded region in the top panel
of Figure \ref{fig:2012cg_velocity}),
which is marginally consistent with our fits. The lightcurve modeling from the
observed $B$, $V$, and $R$ measurements use a distance modulus of
30.9. The summary of our results from the photometric data are presented in
Figure \ref{fig:2012cg}.

\subsection{SN 2009ig}

The velocities and photometry for SN 2009ig are presented in
\citet{Foleyetal2012}. The time of $B$-band peak is at ${\rm JD}~2455080.54$,
and the distance modulus is $32.6$. The evolution of the
Si II $\lambda6355$ absorption feature is a little more complicated
in this case and deserves some discussion. At early times (earlier
than 12 days before $B$-band peak), Si II appears to only have a
high velocity component, and a low velocity component grows to be
more prominent later. We take the low velocity component as
indicative of the photosphere, but use both the high and low
velocity components when fitting the $v\propto t^{-0.22}$ power law.
Data taken when the features overlap could potentially bias the fit
due to blending, but we did not find that it has an adverse impact
on our fits.

\begin{figure}
\epsscale{1.2}
\plotone{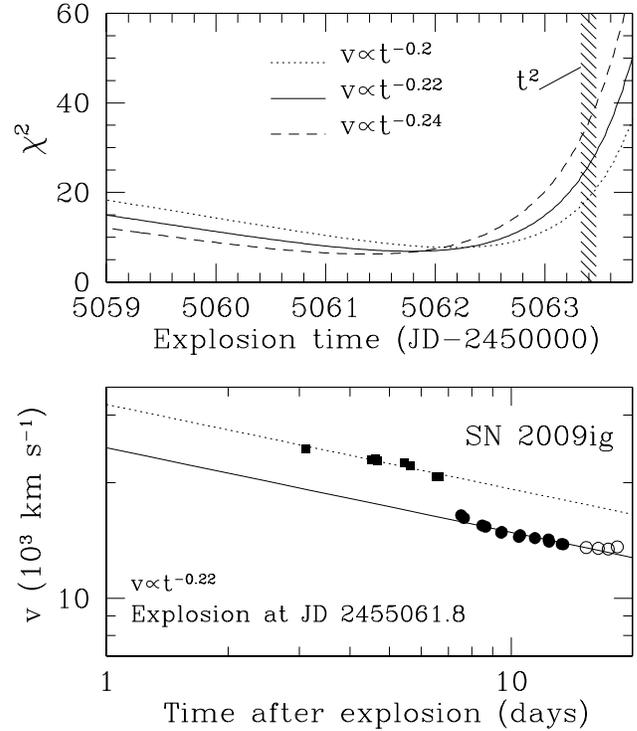}
\caption{The same as Figure \ref{fig:2011fe_velocity}, but for SN 2009ig. Open circles indicate data that was not used for the fit because they are too close to peak. Although an explosion time of ${\rm JD}~2455061.8$ is favored, the constraints are not as strong as for the other SNe.}
\label{fig:2009ig_velocity}
\epsscale{1.0}
\end{figure}

In Figure \ref{fig:2009ig_velocity} we summarize the velocity
fitting. Only high and low velocity Si II are used in this case. High-velocity Ca~II~H\&K absorption features
may be blended with Si II $\lambda4130$, and are not presented by \citet{Foleyetal2012}.
The best fit time of explosion is at
${\rm JD}~2455061.8$. In comparison, using a $t^2$ rise
 \citet{Foleyetal2012} infer an explosion time ${\rm JD}~2455063.4\pm0.07$.
Although SN 2009ig has the least constraining fits of any of the SNe,
this later explosion time seems difficult to reconcile with the
velocity evolution unless $v(t)$ is a much shallower power law with
time than that expected from theory. In Figure \ref{fig:2009ig} we plot the best fit lightcurve
properties.

\subsection{Comparing and Contrasting Events}
\label{sec:compare}

\begin{figure}
\epsscale{1.2}
\plotone{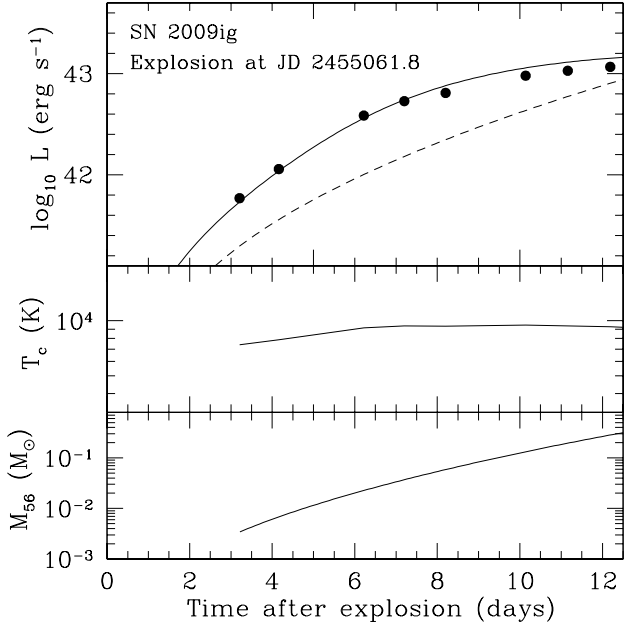}
\caption{The same as Figure \ref{fig:2011fe}, but for SN 2009ig.}
\label{fig:2009ig}
\epsscale{1.0}
\end{figure}

\begin{figure}
\epsscale{1.2}
\plotone{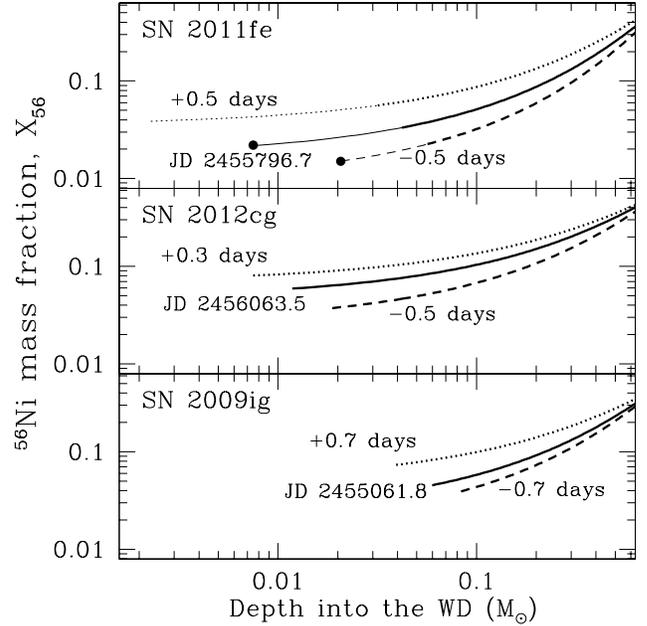}
\caption{Inferred distribution of $^{56}$Ni as a function of depth in the WD. In each case we compare multiple explosion times, with the solid lines indicating the value preferred by fitting $v\propto t^{-0.22}$. The depth into the star is assumed to scale as $\Delta M_{\rm diff}\propto t^{1.76}$ with a normalization of $\Delta M_{\rm diff}=1.4M_\odot$ at lightcurve peak. For SN 2011fe (top panel), the thick curves correspond to the constraints from the observations by \citet{Vinkoetal2012}, and the thin curves correspond to the observations by \citet{Nugentetal2011}. The filled circles indicate the shallowest allowed deposits of $^{56}$Ni so as not to overshoot the upper limit presented by \citet{Bloometal2012}.}
\label{fig:ni56}
\epsscale{1.0}
\end{figure}

In Figure \ref{fig:ni56} we plot the distributions of $^{56}$Ni
inferred for the three SNe Ia modeled above. In each case multiple
values for the explosion time are considered to demonstrate how
inferences on $X_{56}$ change with this parameter.
The solid lines in each panel indicate the
preferred explosion time. For SN 2011fe (top panel), thick
lines show the distribution covered by the photometric observations
of \citet{Vinkoetal2012} and thin lines show the distribution
inferred by the earlier observations by \citet{Nugentetal2011}. This shows that since $\Delta M_{\rm
diff}\propto t^{1.76}$, having observations just a day or two
earlier can probe much shallower regions of the ejecta. For the
preferred explosion time, $X_{56}\approx
2\times10^{-2}$ at a depth of $\Delta M_{\rm diff}\approx
10^{-2}M_\odot$. These results are roughly consistent with
models presented by \citet{Piro2012}, which assumed a similar
explosion time but did not include the diffusive tail. As
discussed in \S \ref{sec:2010fe radius}, the upper limit on the
luminosity at early times implies that there must be a cut-off in the
$^{56}$Ni distribution for some explosion times.
These shallowest $^{56}$Ni depths are indicated by filled circles in the top panel of Figure \ref{fig:ni56} (although not mentioned in \S \ref{sec:2010fe radius}, for the $-0.5\,{\rm days}$ curve, $^{56}$Ni cannot be shallower than the depth of the diffusion wave at $1.7\,{\rm days}$ after the explosion).

The $^{56}$Ni distributions in SN 2012cg and SN 2009ig are fairly similar to SN 2011fe over similar depths. The main difference is that SN 2012cg shows somewhat more $^{56}$Ni around a range of $\Delta M_{\rm diff}\approx10^{-2}-10^{-1}M_\odot$. Does this imply that SN 2012cg has more shallow burning products? Analysis of the spectra indicate that SN 2011fe has considerably more unburned carbon at shallow depths than SN 2012cg \citep{Parrentetal2012}, which is at least consistent with this hypothesis.

SN 2009ig also has a number of differences that are worth discussing. The Si II velocities at \mbox{$\approx10\,{\rm days}$} past explosion are considerably higher in this event than either SN 2011fe or SN 2012cg. If this indicates a difference in the actual explosion energy, then using $v_{\rm ph}\propto E^{0.39}$ \citep{PiroNakar2013} argues that SN 2009ig was a factor of $\approx2$ more energetic than the other two events. Such an explanation seems difficult to reconcile with the peak luminosity of SN 2009ig, which is fairly standard for SNe Ia. Another attractive possibility is that the large velocities are due to an asymmetric explosion that is directed more toward the observer \citep{Maedaetal2010a}. For such larger velocities, there is more expansion and a generally cooler SN, as can be seen by the $T_c$ presented in the middle panel of Figure \ref{fig:2009ig}. \citet{Foleyetal2012} note that SN 2009ig is considerably redder in the UV at early times in comparison to other SNe Ia and typical templates. Is this just due to the larger velocities? Another possibility is that these colors are due iron-peak elements near the surface, which again would be consistent with an explosion directed toward the observer. The mass fraction of $^{56}$Ni for SN 2009ig is fairly similar to the other SNe at a depth of $\approx0.1M_\odot$, and data is not available early enough to probe shallower regions.

Although the many differences found for SN 2009ig are tantalizing, we emphasize that these conclusions all hinge on our assumption that roughly $v\propto t^{-0.22}$. If for some reason the velocity profile of SN 2009ig is different than the other two SNe, then these conclusions must be revised. On the other hand, if the velocity profile {\em is} significantly different in this case, that might be interesting in and of itself. If the supernova is asymmetric, it also limits the applicability of our models, which assume spherical symmetry, in assessing the properties of this event. Properties we infer, such as the $^{56}$Ni, distribution maybe then reflect some sort of angle-averaged property of the ejecta rather than directly measuring the ejecta profiles. Future numerical work should explore how well the correlations we discuss (between velocity, temperature, and so on) still hold for asymmetric explosions, and as a function of viewing angle.

\subsection{Progenitor Models}
\label{sec:progenitors}

For all three SNe we study, $^{56}$Ni must be present at least \mbox{$\approx0.1M_\odot$} from the WD surface, and as shallow as $\approx10^{-2}M_\odot$ from the surface for SN 2011fe and SN 2012cg (see Figure \ref{fig:ni56}). It is therefore worth discussing the implications for progenitor models and the character of the explosive burning.

As a comparison, \citet{Hachingeretal2013} performed detailed UV/optical spectral modeling of SN 2010jn. From this analysis they also infer iron-group elements near the surface. DDT models can produce $^{56}$Ni near the WD surface \citep[e.g.,][]{Iwamotoetal1999}, but to get radioactive material as shallow as $\approx10^{-2}M_\odot$ may require a strongly mixed, off-center deflagration \citep{Maedaetal2010b}. In DDT models with many ignition points that have fairly stratified ashes, radioactive elements are not present near the surface. A gravitationally confined detonation also produces iron-peak elements near the surface when a bubble rises and breaks \citep{mea09}.

Another interesting scenario that may produce shallow radioactive heating is the explosive ignition of a helium shell in a double-detonation. The depth and amount of $^{56}$Ni we infer is not dissimilar to the helium shell masses needed for detonation and the total amount of radioactive material found for such events \citep{sb09,fin10}. The main problem with such models is that if iron-peak elements are too abundant, they tend to produce colors that are too red and spectra that are inconsistent with normal SNe Ia \citep{kro10,Simetal2012}. But if the helium burns in a lateral detonation which does not process the fuel as completely to iron-peak elements \citep{Townsleyetal2012}, this may overcome some of the difficulties double-detonation models have in reproducing observed SNe Ia.

\section{Is a \lowercase{$t^2$} Rise Special?}\label{sec:t^2}

A common practice with recent SN Ia observations is to determine
the time of explosion by fitting the rising luminosity (often in a
single band) with a $t^2$ curve
\citep{Nugentetal2011,mb12,Foleyetal2012,Silvermanetal2012}. Studies
of composite lightcurves formed from stacking many SNe, which
allow the power-law index to vary, find power-law indices of $1.8\pm0.2$ \citep{con06},
$1.8^{+0.23}_{-0.18}$ \citep{hay10}, and $2.20^{+0.27}_{-0.19}$
\citep{gan11}. This begs the question, is $t^2$ (or any
power law) fundamental, and if not, what is the origin of these
results?

Our discussion in \S \ref{sec:theory} shows that a priori a
power-law luminosity rise is not generally expected. The luminosity
is driven by a combination of two factors: (i) the diffusion wave
propagation, $\Delta M_{\rm diff}(t)$, and (ii) the distribution of
$^{56}$Ni fraction, $X_{56}$. The exposed mass does indeed evolve as
a power law, with \citep{Piro2012}
\be
	\Delta M_{\rm diff}(t)\propto t^{2(1+1/n)/(1+1/n+\beta)},
	\label{eq:mdiffpowerlaw}
\ee
where $n$ is the polytropic index and $\beta$ is the power-law index of
the velocity gradient. For $n=3$ and $\beta=0.186$ \citep{Sakurai1960}, this results
in $\Delta M_{\rm diff} \propto t^{1.76}$ (as in eq.~[\ref{eq:mdiff}]).
In contrast, the
$^{56}$Ni distribution is not well constrained by theory and may, in
principle, vary in many ways. A power-law rise of the bolometric
luminosity is expected only if the $^{56}$Ni fraction evolves as a
power law as well, namely $X_{56} \propto t^\alpha$. In this case
the bolometric luminosity evolves as $L \propto t^{1.76+\alpha}$ and
since the photospheric radius is roughly $\propto t^{0.78}$ \citep{PiroNakar2013}, the
observed temperature evolves roughly as $T_c \propto
t^{(0.2+\alpha)/4}$. This result was obtained by
\citet{Piro2012} when the diffusive tail was not included, and
we find that it still holds with the more detailed analysis presented in \S \ref{sec:theory}.
Since we do not expect $X_{56}\propto t^{0.24}$, our conclusion is that
a $t^2$ rise (bolometric or in a single band) is
probably not a generic property of SNe~Ia. We also do not
expect the rise to follow exactly any other power law. Moreover,
since most explosion models predict a sharp decrease of $X_{56}$ in
the outermost layers of the ejecta, the lightcurve is expected to
rise exponentially (due to diffusive tail contribution) at very
early times. How early this exponential phase take place depends on
the depth of the shallowest $^{56}$Ni deposit.

What is then the explanation of the fact that analysis of large SNe
samples are found to be consistent with a power-law rise with
indices in the range $\approx1.8-2.2$? It is probably a combination of two
things. First, the unknown explosion time enables a
reasonable fit even if the lightcurve is not exactly a power law.
Second, in the depth range explored by most of these SNe
rising phases, the  $X_{56}$ is not varying by a large amount.
This is because the first observation of most SNe~Ia take place
only a few days to a week after the explosion, so that the
$^{56}$Ni is distributed roughly uniformly or slowly increasing with depth.

\begin{figure}
\epsscale{1.2}
\plotone{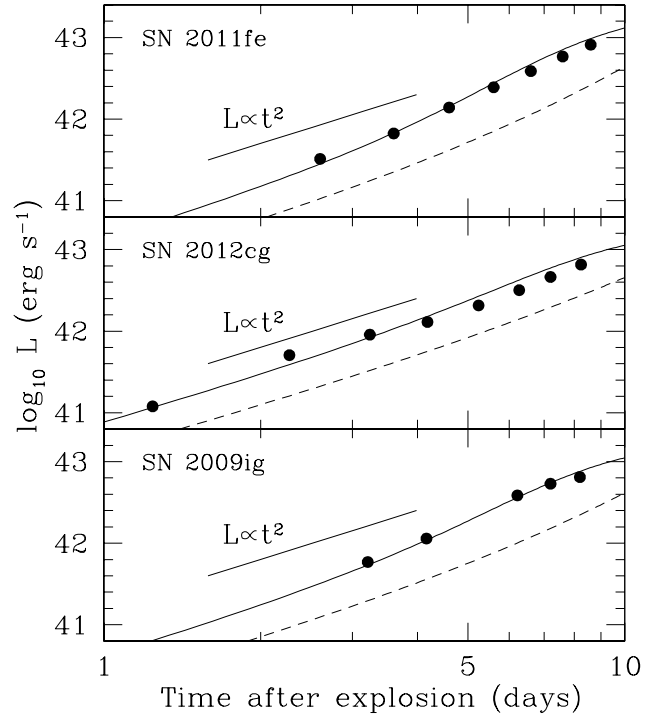}
\caption{The bolometric lightcurves for each of the three SNe (from the top panels of Figures \ref{fig:2011fe}, \ref{fig:2012cg}, and \ref{fig:2009ig}), but in this case plotted with logarithmic axes to emphasize power-law behavior. Again the dashed lines are $L_{56}$.}
\label{fig:t2}
\epsscale{1.0}
\end{figure}

In Figure \ref{fig:t2}, we plot the bolometric lightcurves and fits
for the three SNe we have been studying on a logarithmic scale to emphasize power-law
dependencies. This shows that the lightcurves are not rising exactly
as power laws, but that power law fits can provide a reasonable
description of the data (although for SNe 2001fe the rise
is found to be slightly faster than $t^2$). This is because in all three of these
SNe the $X_{56}$ is rising rather gradually over the depth range
probed by the observations.

There have been some attempts to explain why there should be a $t^2$ rise,
but none of these provide arguments that are expected to hold in detail.
The most simplified explanation is a fixed color temperature with a
radius that increases linearly with time \citep{rie99}. This model
does not explain why the temperature should be constant, and more
importantly, in real SNe the color temperature does typically vary
with time. A more fundamental explanation for a $t^2$ rise is given
by \citet{Arnett1982}, which
considers radioactive heating with thermal diffusion \citep[also see the Supplementary Information of ][]{Nugentetal2011}.
This model makes two explicit approximations: (i) it ignores the
velocity gradient, obtaining $\Delta M_{\rm diff} \propto t^2$
(basically setting $\beta=0$ in eq. [\ref{eq:mdiffpowerlaw}]), and (ii) it assumes
that $X_{56}$ is constant. Together these factors result in a $t^2$ rise, but
only for assumptions that are not realistic.

To conclude, we expect the early rise to depend on the particular
physical conditions in any given event and thus to possibly vary
from one SN to another. It will be important to test this
hypothesis in the future by building bolometric lightcurves from
observations to infer just how much diversity there really is.
Detailed numerical calculations of the rise will also be useful for
understanding how much the early luminosity can change depending on
composition and radiative transfer effects. Whatever the results
are, extrapolating a lightcurve back in time with $t^2$ is not a
reliable method for inferring the explosion time.

\section{Conclusions and Discussion}
\label{sec:conclusion}

Using early observations of three SNe Ia, and assuming that the absorption feature velocities evolve as $v\propto t^{-0.22}$, we constrained the explosion times and shallow distributions of $^{56}$Ni. We then used these findings to revisit the radius constraints on the progenitor of SN 2011fe (in \S \ref{sec:2010fe radius}), and discuss the $t^2$ rise that is reported for many SNe Ia (in \S \ref{sec:t^2}). Using such methods, we are only able to constrain the time of explosion to roughly $\pm0.5\,{\rm days}$ and the corresponding $^{56}$Ni  mass fraction at a given depth to within a factor of roughly $\pm3$. Nevertheless, it is difficult to avoid the fact that $^{56}$Ni must be present at relatively shallow depths ($\sim10^{-2}\,M_\odot$ from the WD surface), even if in very small amounts ($X_{56}\sim10^{-2}$).

SN 2011fe and SN 2012cg appear very similar in most respects, including the rise time, $^{56}$Ni distribution, and energetics. The main difference is that SN 2012cg has a slightly larger amount of shallow $^{56}$Ni. SN 2009ig is somewhat different than the other two SNe. Although its $^{56}$Ni distribution over the same depths probed in SN 2011fe and SN 2012cg are fairly similar, it has higher velocities at any given time and its best fit time of explosion has the largest discrepancy with previous estimates ($\approx1.6\,{\rm days}$ earlier). This is curious because the peak luminosity of SN 2009ig is fairly normal in comparison to the other SNe, and thus the amount of $^{56}$Ni and the energetics should be similar. One possible solution is if SN 2009ig is asymmetric with higher velocities directed toward the observer \citep{Maedaetal2010a}. Unfortunately our results on SN 2009ig are somewhat tentative because it has the least constrained time of explosion. This is because the low velocity Si II is only seen relatively late, and thus has a rather flat evolution with time. Hopefully our work inspires more detailed modeling of SN 2009ig in the future to test our conclusions.

These comparisons show how important it is to have the earliest observations possible. Out the events we consider, SN~2011fe is the best constrained because it shows the largest velocity gradients. Just a few velocity measurements very early in the lightcurve can be more helpful in determining the explosion time than having many measurements at later times. Furthermore, since $\Delta M_{\rm diff}\propto t^{1.76}$, having observations only a day or two earlier probe much shallower depths in the ejecta. Although not discussed much here, having one or two early spectra that can be used for modeling the surface temperature can also provide tight constraints on the explosion time \citep{PiroNakar2013}.

With just these three events, we are already beginning to see correlations between the various features that determine the early lightcurve rise. In the future, studies should look for connections between the early rise and a larger range of properties, such as the late nebular features or the characteristics of the host galaxies. It will also be useful to compare spectral modeling methods for measuring surface abundances \citep[like in][]{Hachingeretal2013} with the techniques we present here.  If used together, they may be more constraining on the nature of the progenitors and the details of the explosive burning. Finally, it would worth exploring the early lightcurves of non-standard SNe Ia, like SN 2002cx \citep{Lietal2003,Foleyetal2013}. Such studies will be important for fully utilizing the observations available  in this new era of early detections of exploding WDs.

\acknowledgments
We thank Ryan Foley, Mohan Ganeshalingam, and Jeffrey Silverman for assistance with assessing data and discussing observations. We also thank Federica Bianco, Ryan Chornock, Luc Dessart, Stephan Hachinger, Keiichi Maeda, Peter Nugent, and Re'em Sari for helpful discussions or comments on previous drafts. ALP was supported through NSF grants AST-1212170, PHY-1151197, and PHY-1068881, NASA ATP grant NNX11AC37G, and by the Sherman Fairchild Foundation. EN was partially supported by an ERC starting grant (GRB-SN 279369) and by  the I-CORE Program of the Planning and Budgeting Committee and the ISF (1829/12).

\end{document}